\documentclass[sigconf]{acmart}

\usepackage{multirow}
\usepackage{makecell}
\usepackage{ulem}

\usepackage{tabularx}
\usepackage{booktabs}

\AtBeginDocument{%
  }

\copyrightyear{2025}
\acmYear{2025}
\setcopyright{cc}
\setcctype{by}
\acmConference[ASSETS '25]{The 27th International ACM SIGACCESS Conference on Computers and Accessibility}{October 26--29, 2025}{Denver, CO, USA}
\acmBooktitle{The 27th International ACM SIGACCESS Conference on Computers and Accessibility (ASSETS '25), October 26--29, 2025, Denver, CO, USA}
\acmDOI{10.1145/3663547.3746401}
\acmISBN{979-8-4007-0676-9/2025/10}

\newcommand{\system}{Task Mode}
\newcommand{\code}[1]{\texttt{\small#1}}

\begin{document}

\title{\system{}: Dynamic Filtering for Task-Specific\\ Web Navigation using LLMs}

\author{Ananya Gubbi Mohanbabu}
\affiliation{%
  \institution{The University of Texas at Austin}
  \city{Austin}
  \country{USA}}
\orcid{0009-0002-1014-4385}
\email{ananyagm@utexas.edu}

\author{Yotam Sechayk}
\affiliation{%
  \institution{The University of Tokyo}
  \city{Tokyo}
  \country{Japan}}
\orcid{0009-0002-5286-0080}
\email{sechayk-yotam@g.ecc.u-tokyo.ac.jp}

\author{Amy Pavel}
\affiliation{%
  \institution{University of California, Berkeley}
  \city{Berkeley}
  \country{USA}}
\orcid{0000-0002-3908-4366}
\email{amypavel@eecs.berkeley.edu}

\begin{abstract}
    Modern web interfaces are unnecessarily complex to use as they overwhelm users with excess text and visuals unrelated to their current goals. 
    Such interfaces can particularly impact screen reader users (SRUs), who may need to navigate content sequentially and thus spend minutes traversing irrelevant elements compared to vision users (VUs) who visually skim in seconds. 
    We present \system{}, a system that dynamically filters web content based on user-specified goals using large language models to identify and prioritize relevant elements while minimizing distractions. 
    Our approach preserves page structure while offering multiple viewing modes tailored to different access needs. 
    Our user study with 12 participants (6 VUs, 6 SRUs) demonstrates that our approach halved task completion time for SRUs while maintaining performance for VUs, decreasing the completion time gap between groups from 2x to 1.2x. 11 of 12 participants wanted to use \system{} in the future, reporting that \system{} supported completing tasks with less effort and fewer distractions. 
    This work demonstrates how designing new interactions simultaneously for visual and non-visual access can reduce rather than reinforce accessibility disparities in future technology created by researchers and practitioners.

\end{abstract}



\maketitle

\section{Introduction}
Billions of people use the web daily to access information, connect with others, and complete essential tasks from purchasing groceries to making medical appointments~\cite{ITU2024MDD}. 
However, web interfaces add unnecessary friction to these everyday tasks.
When a person visits a website to accomplish their goal (\textit{e.g.}, finding a knit sweater on eBay), they must navigate complex layouts designed to simultaneously support a wide variety of user aims, and avoid distracting elements that divert attention from their goal (\textit{e.g.}, banner ads, pop-ups, irrelevant recommendations)~\cite{gray2018dark, moacdieh2017using, monge2023defining}.

Web interfaces present usability issues for everyone, but they impact users differently depending on how they access the web. 
People who use vision to access the web, or vision users (VUs), take around five seconds to identify if a page is relevant then they visually skim for content related their goal~\cite{michailidou2008investigating}, a process that becomes time-consuming and cognitively demanding when the visual design guides their attention towards irrelevant elements (\textit{e.g.}, a bright red banner with holiday deals)~\cite{moacdieh2017using}.
Such visual distractions can be particularly impactful for VUs prone to losing focus on their task.
Blind and low vision people who use screen readers to access the web, or Screen Reader Users (SRUs), can avoid visual distractions but screen readers present information sequentially such that it takes extra time to move past irrelevant content~\cite{bigham2007webinsitu}. 
As interfaces are designed to optimize for visual skimming rather than sequential access, SRUs face persistent usability disparities especially on complex web interfaces. For example, while VUs can assess relevance in 5 seconds, it can take SRUs anywhere from ten seconds to three minutes to just reach the main content of a single page~\cite{takagi2003accessibility}. 
While the disparities on a single interface are small, they quickly multiply as users move through many pages to complete their goal. 
Addressing such disparities requires moving beyond accessibility~\cite{wcag} to close the usability gap.

Prior work has aimed to make websites easier to use for SRUs~\cite{harper2007sadie,bigham2007webinsitu,gubbi2024context,yu2025cluttered} and VUs~\cite{AdBlock2025,lee2025purpose,Apple2024DistractionControl}. For SRUs, prior research explored making websites accessible on-demand by automatically addressing accessibility issues~\cite{harper2007sadie,bigham2007webinsitu,gubbi2024context} or reformatting the webpage to be easier to browse (\textit{e.g.}, by improving the header hierarchy)~\cite{yu2025cluttered}. 
Such tools can support efficient web access, but they are limited in potential efficiency gains as they maintain all content on the page whether or not it is related to user goals.
For VUs and SRUs, other research and commercial tools let users remove certain types of web elements altogether~\cite{AdBlock2025,Apple2024DistractionControl}. For example, ad blockers detect and hide ads~\cite{AdBlock2025}, Lee et al.'s Purpose Mode hides attention capture patterns such as red notification indicators~\cite{lee2025purpose}, and Apple's Distraction Control~\cite{Apple2024DistractionControl} lets users hide elements of their choice (\textit{e.g.}, a recommendations pane). 
Such tools can reduce distractions and improve efficiency. However, the usefulness of a given web element (\textit{e.g.,} a product listing or menu item) often depends on its relevance to the users' goal, not just the element’s type.
Large pretrained models including large language models (LLMs) and Multimodal Large Language Models (MLLMs) have potential to determine what web elements are relevant to a user goal. 
The key insight of our work is that there exists an opportunity to support users in semantically filtering webpage content according to their goal to increase efficiency for everyone.

In this work, we contribute \system{}, a system to improve the usability of websites by dynamically adapting websites to fit users' stated goals as they use the web.
With \system{}, users can enter their goal into a web extension (\textit{e.g.}, \textit{``buy a laptop under \$500 that isn't a Chromebook''}),
then \system{} uses an LLM-based pipeline to surface relevant elements and hide irrelevant elements.
For example, if the user visits the eBay~\cite{ebay} homepage to buy their laptop, 
our extension will surface navigation elements (\textit{e.g.}, the ``Electronics'' menu item, the search bar, a flash deal for laptops) related to the high level task and hide or de-emphasize elements that are unrelated (\textit{e.g.}, the ``Motors'', ``Home \& Garden'', and ``Toys'' menu items, and the front page recommendations for shoes and clothes). The \system{} persists as users move onto other websites (Amazon~\cite{amazon}, Best Buy~\cite{bestbuy}) and into particular item pages (\textit{e.g.}, surfacing the ``Add to Cart'' button and item details while suppressing the ``Careers'' page). 
Our system lets users update their task as they go, select their relevance threshold to surface or suppress more elements, and select view options to meet their access needs including removing irrelevant elements from the page (particularly useful for SRUs) and making more relevant elements more visually salient (particularly useful to VUs).
With \system{}, users have the agency to choose whether they want to minimize distraction or maximize exploration.

We prototyped \system{} as an LLM-powered web browsing extension, and evaluated \system{} in a user study (N=12) with six VUs and six SRUs who frequently used the web to accomplish daily tasks. 
The study session consisted of two stages: a controlled comparison to manual webpage navigation, and an open-ended task where users tried \system{} for their own tasks. 
In the controlled comparison, SRUs reduced their task completion time from 211 ($\sigma$ = 49.1) seconds with traditional techniques to  102 ($\sigma$ = 26.2) seconds with ($p < 0.05$), while VUs task completion time changed from 107 ($\sigma$ = 39.9) seconds with traditional techniques to 84.4 ($\sigma$ = 33.3) seconds with ($p > 0.05$). 
Thus, the task completion time disparity between SRUs and VUs decreased from a 2x difference with traditional browsing to a 1.2x difference with \system{} --- reducing the usability gap. 
Both SRUs and VUs (11 of 12 total participants) wanted to use \system{} in the future because it lowered effort and distractions as they browsed the web. 

Our work designs, develops, and evaluates a new system for both VUs and SRUs from the beginning. We achieve interactions that both VUs and SRUs find beneficial while being most impactful for SRUs, thus reducing the usability gap. As Human Computer Interaction (HCI) practitioners design new interfaces targeted at optimizing interactions for VUs at an alarming rate (\textit{e.g.}, new AR/VR experiences, canvas-based human-AI interactions), our work suggests the potential to design for both visual and non-visual access at once such that future innovations in the HCI community ameliorate rather than exacerbate the usability gap of technology at large.

\section{Related Work}
As we aim to create task-specific web navigation interaction, our work builds on three complementary research areas: assistive technologies for web navigation that address accessibility barriers for blind and low-vision (BLV) users, customized user interfaces that adapt to users' needs, and web agents that automate tasks.

\subsection{Assistive Technologies for Web Navigation}
Many websites are designed to guide user's attention to locate essential elements based on visual cues (\textit{e.g.}, a large red button that says ``Apply Now!'').
However, on websites the visually salient cues often do not relate to user goals, and sometimes such cues are intentionally used in Attention Capture Damaging Patterns~\cite{lee2025purpose} to distract users from their goals. 
Screen reader users (SRUs) who rely primarily on non-visual navigation strategies are excluded from using visual cues for quick navigation~\cite{chiou2023bagel, silva2024page}. 
Thus, while SRUs avoid visual distractors, it can be challenging for SRUs to build accurate mental models of webpage layouts, as these models are often shaped by visually oriented design conventions they cannot directly access~\cite{chheda2023understanding, baez2022exploring}. These challenges are compounded by persistent accessibility barriers: large-scale audits confirm that web accessibility issues remain pervasive with over 95\% of homepages having detectable WCAG  failures ~\cite{WebAIM, wcag}, including missing form labels, poor color contrast, and absent alt texts in images.

While screen readers provide linear access to web content, SRUs must often sequentially navigate through substantial irrelevant content to reach their goals~\cite{baez2022exploring, prakash2024all, ashok2017web, alves2018automatically, prakash2023autodesc}.
To speed up navigation, SRUs use strategies like heading navigation and landmark skipping to jump between key elements instead of reading linearly~\cite{WebAIM}. These strategies reduce sequential access burden, but are limited since many webpages lack proper use of semantic HTML tags or have inconsistent structures, making it hard for SRUs to quickly find relevant information.
For instance, a visually evident section like \textit{``featured products''} may require extensive sequential navigation, including irrelevant content such as advertisements or page headers, before reaching the desired information. This problem worsens on websites with dense layout where elements are inconsistently arranged or mislabeled leading to disorientation, fatigue, and difficulty forming mental models of page structures~\cite{silva2024page, yu2025cluttered, chiou2023bagel, chheda2023understanding, prakash2023autodesc}. 
Additionally, commercial tools like JAWS Smart Glance ~\cite{Smart_Glance_Feature_Jaws} can help users find visually salient elements such as bold or large text. However, they lack awareness of user intent and can miss task-relevant content that is not visually prominent.

Prior work has proposed several strategies to improve web navigation for BLV users including content restructuring ~\cite{yu2025llm, yu2025cluttered, silva2024page}, automated summarization ~\cite{alves2018automatically, baez2022exploring}, inserting semantic markup ~\cite{bigham2007webinsitu, harper2007sadie}, and semantic labeling ~\cite{ashok2017web, wang2021revamp}. Other works support customization by letting users create overlays ~\cite{todi2018familiarisation} or insert navigation points for easier skimming ~\cite{baez2022exploring}. Content restructuring approaches ~\cite{yu2025llm, yu2025cluttered, silva2024page}, simplifies layout by modifying the DOM, but may obscure intent or disrupt expected flows for SRUs. Other approaches inject helpful accessibility annotations automatically, such as adding ARIA roles and landmarks to improve screen reader navigation ~\cite{brown2013dynamic}. All of these prior approaches improve webpage accessibility, but users must still navigate content irrelevant to their goals. 

Thus, rather than restructuring or summarizing the webpage, we preserve the original web layout and augment the user's browsing experience by removing task-relevant elements for SRUs to reduce irrelevant navigation steps.
One prior study explored manually altering a mock webpage to remove redundant and irrelevant content for a given task and found that this alteration improved task completion time for SRUs \cite{giraud2018web}. However, no prior work has explored the feasibility and efficacy of such an approach when automatically determining element usefulness on highly complex real-world websites. Our work thus contributes the design, implementation, and evaluation of a system to surface task-relevant elements. 

\subsection{Customizable User Interfaces}
As one-size-fits-all interfaces may not be the best fit for individual users and their goals, customizable user interfaces have been a longstanding goal in Human-Computer Interaction (HCI).
Prior work has extensively explored methods for customizing and optimizing user interfaces (UIs) across various computing platforms, including web~\cite{zhou2010optimizing, todi2018familiarisation}, mixed reality~\cite{lindlbauer2019, cheng2021}, desktop~\cite{gajos2004supple, 10.1145/2702123.2702149}, and mobile devices~\cite{zhou2010optimizing}. For instance, SUPPLE automatically generates adaptive interfaces to optimize user interaction based on device constraints and user-specific preferences~\cite{gajos2004supple}. Other research has specifically targeted adaptation for user abilities, dynamically adjusting interfaces to individual users' physical or cognitive capabilities~\cite{gajos2008improving, gajos2007automatically, gajos2006automatically, motti2013, sarcar2018}. Other existing approaches have considered real-time adaptation to repetitive tasks, improving efficiency during frequently performed activities~\cite{gajos2006designspace}.
While building upon these insights, our approach does not require predefined mathematical models of user tasks, which allows us to generalize to new unseen tasks and adapt the user navigation experience real-time as their goals evolve. We also preserve original webpage structures to retain user's original mental map of where elements are located on specific web pages.

Additionally, recent tools like Purpose Mode~\cite{lee2025purpose} and Apple’s Distraction Control~\cite{distractioncontrol} allow users to suppress distracting elements across websites. However, these tools rely on manual element selection, and we aim to automatically detect which elements are relevant to the user's tasks and which are not.

While much of the literature focuses on optimizing interfaces for single-page or single-session tasks, users often pursue longer-horizon goals that span multiple pages or applications~\cite{ma2023browsing, liu2019task}. Malleable interface systems have attempted to support such workflows by enabling users to reshape or recompose interfaces~\cite{klokmose2015webstrates, min2023demonstration, min2025malleable, cao2025generative}. For instance, MasonView~\cite{min2023demonstration} allows users to organize content using viewport-based layouts, while Malleable Overview Interfaces~\cite{min2025malleable} generate page summaries to provide higher-level overviews. However, these systems abstract or restructure the original layout and do not evaluate the task-relevance of individual elements within a web page. Our approach preserves the original webpage layout while dynamically filtering or de-emphasizing elements in real-time, based on user-defined tasks to support both VUs and SRUs in accomplishing more complex and multi-step goals.

\subsection{Web Agents for Task Automation}

Both industry~\cite{operator, computeruse, browseruse, nanobrowser, taxyai} and academic researh~\cite{huq2025cowpilot, pan2024webcanvas, lu2024weblinx, shi2017world, liu2018reinforcement, zheng2024gpt, he2024webvoyager} have explored the use of web agents to automate tasks via web navigation. These agents perform interface actions such as clicking, typing, and navigating, typically driven by user-defined goals.
Early approaches used reinforcement learning (RL) to learn interaction policies~\cite{liu2018reinforcement, shi2017world}, but these agents often struggled to generalize across unfamiliar or dynamic webpages~\cite{humphreys2022data}. Nonetheless, task-specific agents specialized fine-tuned for specific applications have shown promise for SRUs ~\cite{kodandaram2024, ashok2015capti, zhong2014justspeak}.

More recent agents incorporate LLMs to interpret user goals and reason about actions directly from webpage content~\cite{zheng2024gpt, he2024webvoyager, lu2024weblinx}, often using visual context from screenshots to leverage vision-language capabilities ~\cite{lai2024autowebglm, hong2024cogagent}. 
For instance, WEBLINX enables conversational guidance for completing tasks ~\cite{lu2024weblinx}. However, most agents primarily target actionable interface elements (e.g., buttons, links), neglecting non-clickable yet semantically or visually important content such as headings, images, and text blocks that are essential for complex tasks.
Current agents remain unreliable ~\cite{zhou2023webarena, baechler2024screenai, deng2023mind2web, zheng2024webolympus, kodandaram2024} often requiring users to manually intervene and  correct errors ~\cite{huq2025cowpilot}. For SRUs, recovering from agent failures can be especially burdensome, as it requires navigating between the agent interface and the webpage using a screen reader. Moreover, most agents are designed for fully automated, end-to-end task execution with minimal user input, rather than supporting open-ended tasks with multiple decision points similar to scenarios that frequently arise in real-world web browsing.

Instead of acting autonomously on behalf of the user, our system keeps users in control by dynamically filtering web content according to their goals. Unlike prior agents that focus solely on clickable elements, our approach semantically scores and visualizes the relevance of all webpage elements while preserving the original page layout. This design supports both VUs and SRUs in more general and everyday web navigation tasks.

\section{\system{}}
We present \system{}, an accessible system that optimizes web navigation by dynamically ranking task-specific web content. \system{} helps both VUs and SRUs complete online tasks more efficiently by surfacing elements relevant to their stated goal and suppressing irrelevant information (by de-emphasizing it visually for VUs or hiding it entirely for SRUs) to reduce distraction, effort, and time.

\subsection{Design Goals for Dynamic Filtering of Task-specific Web Elements}

Grounded in prior work on task-based browsing, assistive web navigation, and customizable interfaces, we identify five key design goals to guide the development of a system for dynamically filtering task-relevant web content. Each goal reflects a specific challenge observed in the literature and aims to support effective, efficient, and accessible task completion on the web: \\

\begin{enumerate}
    \item[\textbf{D1.}] \textbf{Prioritize task-relevant elements.} The system should identify and foreground web elements that are directly relevant to the user’s stated task. It must reliably interpret the user’s intent based on a task description and de-prioritize or suppress content that does not contribute to task completion. Prior work has shown that surfacing task-relevant content can significantly improve efficiency for SRUs ~\cite{giraud2018web}. 
    \item[\textbf{D2.}] \textbf{Preserve page structure.} Filtering must retain the overall structure of elements on the webpage to maintain navigational consistency. Disrupting the underlying structure can hinder comprehension and cause confusion, particularly for BLV users who make use of predictable page organization in absence of visual cues ~\cite{yu2025cluttered, Understanding_Consistent_Navigation, yu2025llm}. Thus, the system should adapt content presentation while preserving the user’s mental model of the webpage.
    \item[\textbf{D3.}] \textbf{Maintain persistent task context across navigation.} Prior research ~\cite{baez2022exploring} highlights that users often complete multi-step tasks across multiple pages and prefer systems that remember their goals, supporting our goal that the system must maintain the task context across page transitions. 
    Dynamic filtering should be re-applied in a consistent manner as the user navigates through different content.
    \item[\textbf{D4.}] \textbf{Accommodate evolving task formulations.} As users make progress or encounter new information, their understanding of the task may evolve. The system should therefore support real-time updates to the task description and re-interpret content relevance accordingly. This adaptability aligns with research in goal-driven interaction and iterative query reformulation, which shows that users benefit from systems that can adjust to changing goals without restarting the task flow ~\cite{chen2021towards, sloan2015term}.
    \item[\textbf{D5.}] \textbf{Support personalization of filtering and display parameters.} To address the diverse needs of both VUs and SRUs, the system should allow for customizable filtering thresholds, display modes, and interaction options. Personalization has been shown to enhance both usability and satisfaction in accessibility tools, and is critical for ensuring that filtered views remain meaningful and usable across contexts ~\cite{fritz2019customization,alsayahani2025effects}.
\end{enumerate}

\subsection{User Scenario}
To illustrate the use of \system{} for both VUs and SRUs, we follow \textbf{Mike}, a blind student preparing for the new semester and \textbf{Tara}, a sighted shopper browsing the web for adventurous gear on sale.

\subsubsection{Mike (Screen Reader User)}
Mike uses a screen reader to navigate the web. With the new semester approaching, he needs to apply for financial aid to help cover his tuition. He opens his university’s website and \system{}, sets the rendering style to \textit{only task-specific information visible} with a relevance threshold of 70\%, and inputs his task: \textit{“find financial aid application for computer science program tuition.”} \system{} analyzes the page, filtering out everything unrelated to financial aid applications, tuition and costs, or the search bar. This includes sports event promotions, campus activity calendars, and unrelated academic program listings. 

Almost immediately, Mike’s screen reader reads out “Admission \& Aid” in the main navigation menu and “Financial Aid” in the sub-menu, allowing him to quickly land on the “Financial Aid \& Scholarships” page. Instead of sifting through irrelevant content or clicking multiple links, Mike can now focus on what he needs: application links, important deadlines, and contact information. 
After locating all the relevant materials, Mike files his financial aid application. Shortly afterward, he lowers \system{}’s relevance threshold to 40\% to surface broader scholarship options, including grants or departmental funding, thus getting a comprehensive view of all financial aid opportunities while avoiding unnecessary clutter.

\subsubsection{Tara (Vision User)} With summer approaching, Tara needs to purchase camping gear for an upcoming trip with friends. She visits the backcountry.com homepage where she’s immediately overwhelmed by the abundance of menu options, promotions, and product listings. While she’s tried ad-blockers in the past, she prefers using \system{} for its ability to visually highlight items that match her specific needs. She opens the \system{} chrome extension, selects the lower opacity display mode, and enters her initial goal: \textit{``Find a 3-person tent under \$500 with good weather protection''}. \system{} interprets this goa   l in the context of the page and dims unrelated content. The ``Camping \& Hiking'' category, a blog with a tent cover photo titled ``The Camping Guide'', and a sale on tents appear in full opacity. Important interface elements like the search bar remain clearly visible to help her complete the task.
She quickly spots a well-rated 3-person tent on sale for \$279 (15\% off) and clicks the listing. On the product page, \system{} emphasizes relevant details: product name, price, weight, weather rating, customer reviews mentioning durability, and the ``Add to Cart'' button. Tara saves the option and updates her task with the price \$300 to reflect her now lower price criteria then moves onto other sites to compare options (REI, her local outdoors stores, and Amazon) as \system{} continues to surface information efficiently across web pages. When she reaches Amazon there are many viable options, so she switches to the colored gradients display mode to make it more clear what options best match her criteria. With \system{}, Tara avoids irrelevant listings and recommendations, and makes sense of new sites quickly such that she saves time and cognitive effort during her search.

\section{Prototype System}

\begin{figure*}
    \centering
    \includegraphics[width=7in]{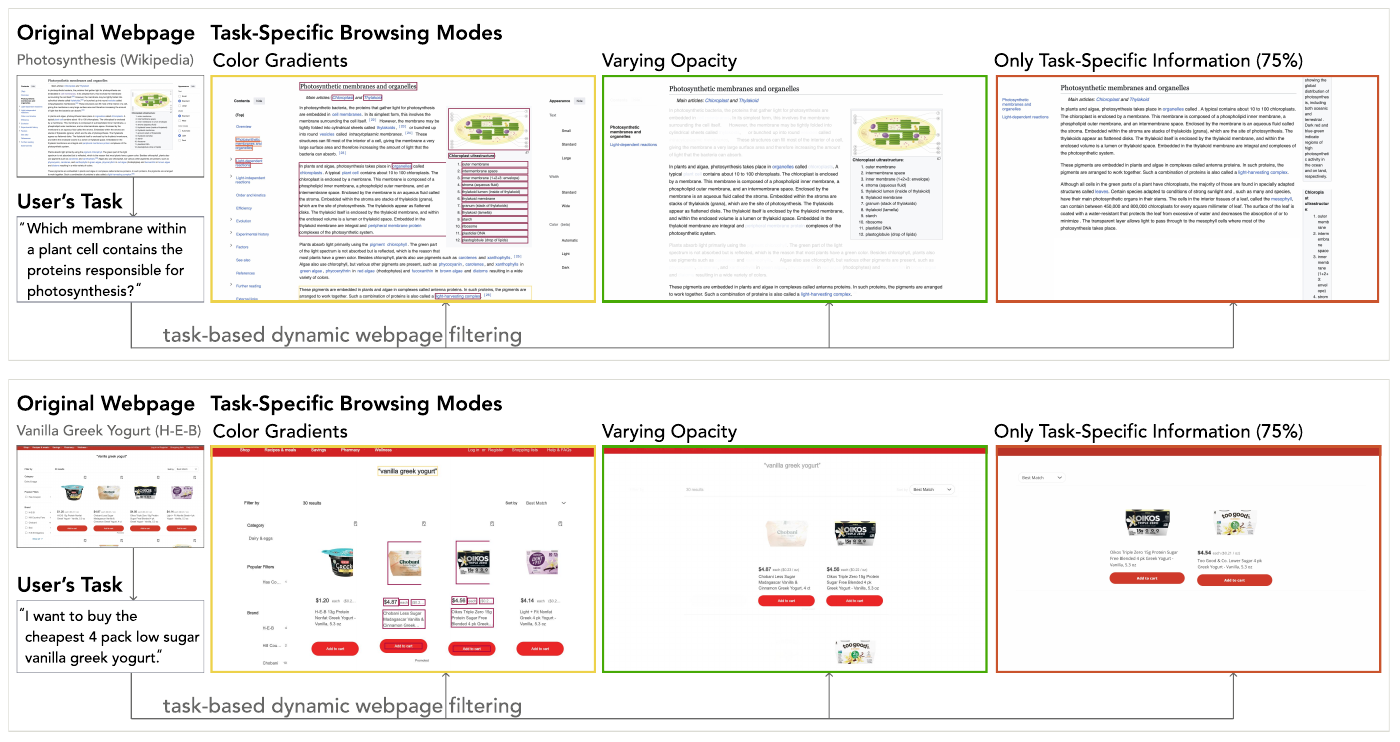}
    \caption{Comparison of task-specific rendering modes. This figure shows the three rendering modes (Color Gradients, Varying Opacity, and Only Task-Specific Information - 75\%) modify two types of webpages based on user tasks. In the top example (Wikipedia), the Color Gradients overlay heatmaps on web elements (texts, images, SVGs) proportional to their task relevance, while Varying Opacity fades out less relevant content, and Only Task-Specific Information (75\%) threshold mode removes low-relevance sections entirely to show only the most pertinent definitions and diagrams. In the bottom example (H-E-B), the Color Gradients highlight product names, prices, and filters that best match the user’s goal, Varying Opacity dims unrelated items and interface elements, and the Only Task-Specific Information (75\%) threshold mode filters out all but the most relevant product listings. Each mode progressively reduces visual noise by suppressing or removing unrelated content, while retaining context to varying degrees.}
    \Description{A figure comparing three filtering styles applied to two webpages. The first row shows a Wikipedia article on photosynthesis with the task: “Which membrane within a plant cell contains the proteins responsible for photosynthesis?” The Color Gradients view adds red and pink overlays on relevant text; the Varying Opacity view fades out surrounding less relevant text; the Only Task-Specific Information view shows only key text and diagrams. The second row shows a grocery website listing yogurt products for the task: “I want to buy the cheapest 4 pack low sugar vanilla Greek yogurt.” Color Gradients highlight product titles and prices; Varying Opacity fades out unrelated items; and the 75\% threshold view displays only matching product cards, removing other page elements.}
    \label{fig:visualization-modes-new}
\end{figure*}

\begin{figure}
    \centering
    \includegraphics[width=2.5in]{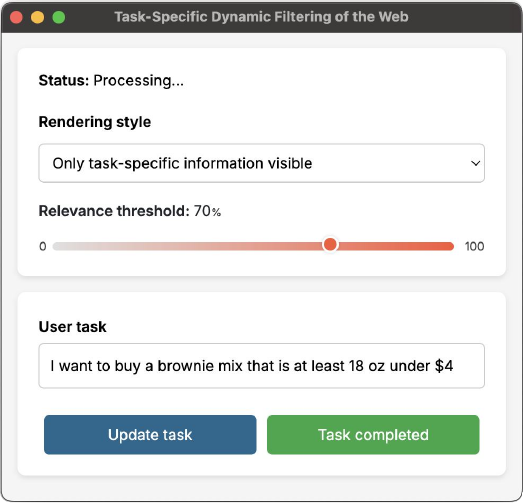}
    \caption{Interface of \system{}. Users specify their task in natural language, and \system{} filters the webpage to show only task-relevant content based on the selected rendering style and adjustable relevance threshold (center). Users can tune the threshold to control filtering strictness and switch rendering modes to include more or less context. The "Update task" and "Task completed" buttons allow users to refine or finalize their task.}
    \Description{A user interface titled “Task-Specific Dynamic Filtering of the Web” is shown. At the top, the status reads “Processing…”. A dropdown labeled “Rendering style” is set to “Only task-specific information visible.” Below it is a horizontal slider labeled “Relevance threshold,” currently set to 70\%. The user task is written in a text box: “I want to buy a brownie mix that is at least 18 oz under \$4.” Below the text box are two buttons: “Update task” in blue and “Task completed” in green. The interface allows users to filter webpage content based on the specified task, rendering style, and relevance threshold.}
    \label{fig:interface}
\end{figure}

\subsection{Interface}
We developed a Chrome Extension (Figure ~\ref{fig:interface}) that users can install and activate on any webpage by clicking the extension’s icon in the browser’s extension bar. Upon activation, the extension opens a dedicated interface window that allows users to add their task description and set their visualization preferences. Once users provide their task description, the extension highlights task-specific elements on the page, allowing them to begin and carry out their task immediately. Users can directly revise their original task description at any time as their information needs evolve. 
The extension also allows users to choose their preferred method of visualizing task-relevant information. We offer three display options: (a) showing all web elements (including both task-relevant and irrelevant content) using color gradients to indicate relevance; (b) displaying all content with gradually reduced opacity based on decreasing relevance to the task; and (c) dynamically filtering and presenting only task-specific information while removing residual irrelevant elements from the page. If users opt to use only task-relevant information, they can dynamically adjust the relevance threshold to control the percentage of task-relevant content displayed. 
The extension also provides real-time system status updates to keep users informed of processing progress. 
For a given task, the extension consistently highlights task-specific web elements on each visited page in the active browser window, thereby supporting the user’s task progression and maintaining consistency within a browsing session. 
Once the user confirms task completion, the extension automatically deactivates.

\subsection{Pipeline}
When the user clicks on the extension icon on a specific webpage where they intend to start carrying out their task, our system initiates a processing pipeline designed to highlight task-relevant content on that page. Specifically, we extract the complete HTML structure of the webpage, capturing both textual elements (such as text, buttons, and links) and visual components (images, SVGs, and iFrames). This extracted data is then processed to identify and filter the content most likely to be relevant to the user’s task. Finally, the identified web elements are highlighted according to the user's selected visualization preference.

\subsubsection{Extracting Webpage Elements.}
Webpages often contain useful information presented through various types of elements, such as textual tags (\code{<p>}, \code{<span>}, and \code{<h1>}) through \code{<h6>}), hyperlinks (\code{<a>}), buttons (\code{<button>}), images (\code{<img>}), icons (\code{<svg>}), and advertisements or embedded content (\code{<iframe>}). Since different types of elements on a webpage may be necessary and relevant for users to complete their tasks, we extract all potentially useful elements by accessing the webpage’s Document Object Model (DOM), excluding non-informative tags (\code{<script>}, \code{<style>}, and \code{<noscript>}). 
We traverse the original DOM structure and construct a duplicate tree for further processing to identify elements related to user task. For each node in the DOM, we assign a unique identifier, extract its tag name and any associated text content, and record its parent and children using their respective IDs. Each DOM node is also initialized with a relevance flag set to false. Visual elements such as \code{<img>}, \code{<svg>}, and \code{<iframe>} are stored and processed separately. For these elements, we extract their source URLs and accompanying metadata, such as the \code{alt} attribute for images, \code{<path>} attribute for SVG and \code{<title>} for iFrame elements. These alternative texts and titles occasionally contain concise descriptions of key visual content and can be useful for assessing the element’s relevance to the user’s stated task.

\subsubsection{Identifying Task-Specific Web Elements.}
Modern large language models (LLMs) exhibit strong capabilities in interpreting user interfaces and reasoning about contextually appropriate next steps in task-based browsing. To assess their effectiveness in identifying task-specific web elements, we evaluated GPT-4o, Claude 3.7 Sonnet, and GPT-o1 using the extracted textual content of webpages in conjunction with corresponding task descriptions. Although GPT-o1 demonstrated marginally higher accuracy in detecting relevant elements (such as correctly identifying item prices in online shopping scenarios), this improvement came at the cost of increased latency and computational overhead. Given that typical webpages contain thousands of elements, maintaining system responsiveness is essential for enabling timely and seamless user interaction. As our goal is to support realistic web usage where users expect immediate access to relevant content, we prioritized responsiveness over incremental gains in accuracy. We selected GPT-4o as the foundation of our pipeline due to its favorable trade-off between processing speed, cost-efficiency, and acceptable accuracy. To further improve the model’s ability to interpret webpage structure and context, we augment its input with additional processing and assign task relevance scores to all potentially useful elements.

\textbf{Processing Textual Information.} To help the model focus on the most contextually relevant parts of a webpage during inference, we assign a \textit{task relevance score} to each text segment using GPT-4o. This score indicates how relevant each webpage element is to specific components of the user’s task (e.g., the item they are searching for or the conditions they specified). These scores help the model identify and focus on the parts of the page most likely to support task completion. Specifically, we first decompose the user's task description into a structured representation with five components using GPT-4o (see A.1 for full prompt),
\begin{enumerate}
    \item \textit{\textbf{Entity}}: the object or subject that the user is searching for.
    \item \textit{\textbf{Constraints}}: preferences or conditions that guide the task.
    \item \textit{\textbf{Actions}}: the likely logical steps the user may take to accomplish the task on a website.
    \item \textit{\textbf{Defaults}}: common elements that users expect to see in the context of their tasks.
    \item \textit{\textbf{Fallbacks}}: alternative content or features that users can rely on for task progression if expected content is unavailable.
\end{enumerate}

We use this structured breakdown of the task to guide GPT-4o in evaluating the relevance of webpage content. Specifically, we provide GPT-4o with a list of all text extracted from the webpage's DOM, along with associated metadata such as HTML tag types, element IDs, and the order in which these elements appear on the page. This structure provides the model with implicit contextual cues, such as grouping, layout proximity, and element hierarchy, which help it reason about relationships between elements.

Importantly, we do not treat elements in isolation. We explicitly instruct the model to consider the broader context in which each element appears as many elements only become meaningful when interpreted in relation to surrounding content. For instance, a price tag like “\$499” may seem irrelevant on its own, but if it is located next to a product labeled “HP Laptop” it becomes highly relevant to a task like “find a laptop under \$500.” Conversely, treating all price tags as equally important regardless of their association with the correct product could mislead the model by inflating the significance of unrelated content. To avoid this, we prompt GPT-4o to consider contextual relationships between elements such as those implied by DOM nesting, shared containers, or visual grouping. This enables the model to infer meaningful clusters of information that more accurately reflect how users interpret a webpage both visually and functionally.

Finally, GPT-4o assigns a task relevance score to each text element on a scale from 0 (completely irrelevant) to 100 (critical to the task). These scores are based on two factors: (1) the element’s direct alignment with one or more components of the task breakdown, and (2) its contextual associations with other relevant elements on the page (see A.2 for full prompt).

\textbf{Processing Visual Information.} We process visual content separately from textual elements, focusing on two key categories: images and SVG icons.
For \textbf{images}, we evaluate if each image is relevant to the task by computing the similarity between the task breakdown and the image's existing alt text with GPT-4o (\textit{Alt Score}, see A.3 for full prompt), and between the task text and the image itself with CLIP~\cite{hessel2021clipscore} (\textit{Image Score}).
The final \textit{task relevance score} for each image is computed as a weighted combination, 0.3 $\cdot$ \textit{Task-Alt Score} + 0.7 $\cdot$ \textit{Task-Image Score}. This weighted scoring emphasizes the similarity between the task breakdown and the image's description as the alt text, when present, often contains task-relevant details that support accurate task to image similarity scores. 
When the image does not have alt text, we use only the \textit{Task-Image Score} such that matching performance may degrade but \system{} maintains its core functionality.
The specific ratio was selected heuristically based on preliminary experiments across multiple task–webpage pairs. 
SVG \textbf{icons} are common on webpages but do not have human-readable descriptions.
The \texttt{<path>} element in SVG is used to define shapes through a series of commands and coordinates. These paths can represent simple lines, curves, or complex structures.
We pass the \texttt{d} attribute of the SVG \texttt{<path>} element as input to GPT-4o, which can infer a semantic label for the icon directly from its path (e.g., search, cart, settings) (see A.4 for full prompt). Once the icon has a label we evaluate its \textit{task relevance score} using the same breakdown of the user's task. Icons that represent actionable steps or convey key information (e.g., a search icon for locating products) are scored based on how well they align with the user’s task objectives (see A.5 for full prompt).

\subsubsection{Relevance-Based Content Visualization}
After scoring each web element for task relevance, we apply different visualization modes to communicate relevance based on user preferences. Each mode highlights relevant content differently, aiming to help users efficiently identify the parts of the page most useful for their task.

\textbf{Showing All Web Elements using Color Gradients.} This mode is designed to support VUs by using color to indicate how relevant each element on the page is to the user's task. It enables users to quickly scan the page and visually prioritize content based on task relevance.
In this mode, we apply a continuous gradient to show task relevance across webpage elements. Specifically, elements with low task relevance are shaded light green, medium-relevance elements appear yellow, and high-relevance elements are marked in red. Each element is scored and color-coded independently, without affecting the appearance of its parent or child elements. Thus, task relevance scores are not propagated through the DOM tree (Figure ~\ref{fig:relevance-propagation}), so hierarchical relationships do not influence the visual encoding. We hide the borders (by setting border opacity to 0) of non-textual container elements such as \code{<div>} and \code{<body>} which typically do not contain visible content themselves to reduce visual clutter and emphasize meaningful content. This ensures that attention is focused on leaf nodes (such as text, images, and SVGs) that are more likely to convey useful task-relevant information.

\textbf{Displaying All Content with Gradually Reduced Opacity based on Decreasing Relevance to the Task.}
This mode visualizes the task relevance of webpage content by adjusting element opacity: elements with higher relevance remain fully opaque, while those with lower relevance gradually fade out. This creates a subtle visual filtering effect that guides VUs attention toward more relevant content while still maintaining the full page layout.

Instead of mapping raw relevance scores directly to opacity values, we first propagate scores within the DOM hierarchy. We propagate scores from child nodes to their parents to prevent relevant content from being unintentionally hidden Figure ~\ref{fig:relevance-propagation}). Specifically, a parent node retains its own relevance score unless one or more of its children has a higher score. In such cases, the higher child score is assigned to the parent. This ensures that containers with meaningful nested content remain visible. Once this propagation is complete, we compute the final opacity values based on the updated scores. This design supports visual skimming by de-emphasizing low-relevance content without altering the overall page structure.

\textbf{Dynamically Filtering and Presenting Only Task-Specific Content.}
This mode primarily supports SRUs by minimizing irrelevant content and displaying only the elements most related to the task.
We assign a task relevance score to each node to determine which elements are displayed. Elements with scores above a predefined threshold are considered relevant. To preserve the overall page structure and ensure a coherent reading experience, relevance scores are propagated upward through the DOM hierarchy: any ancestor of a relevant node (up to the \code{<body>} element) is also retained in the filtered view Figure ~\ref{fig:relevance-propagation}). This ensures that task-specific content is not orphaned or visually disconnected.

Elements with scores below the threshold are completely hidden from both visual rendering and assistive technologies. We set \code{aria-hidden="true"} and disable interaction for these low-relevance nodes. This ensures that SRUs can skip over irrelevant content entirely, allowing them to focus only on elements that are useful for completing their task. This mode performs structural filtering that improves accessible navigation for SRUs, unlike the color gradient or reduced opacity modes that are primarily intended to support VUs by providing visual cues. Additionally, users can dynamically adjust the relevance threshold to control the amount of content displayed, allowing them to access additional context surrounding task-relevant information.

\subsection{Implementation}
We implemented our system as a Google Chrome Extension\footnote{github.com/UT-CS-HCI/task-mode} using JavaScript. We used Python Flask backend with a real-time Firebase ~\cite{firebase2025} database to log all webpage elements with their relevance scores for different webpage-task pairs. Specifically, we use a Firebase real-time database to log the webpage URLs, users' task, webpage elements and their task-specific relevance scores. 

\subsection{Robustness and Compute Cost of \system{}} 
To test the robustness and cost of \system{}, we randomly selected 20 websites from the list of top-100 most visited websites in the US ~\cite{SE_Ranking_2025} (e.g., Amazon, Reddit, Walmart, NYTimes). \system{} successfully parsed and highlighted all interface elements according to their predicted task relevance on 19 of 20 websites. 
One of the websites used ``virtualized infinite scroll'' that reuses a fixed number of components in the DOM by dynamically updating its content as user scrolls and thus while \system{} successfully parsed initial elements, it missed all elements modified while scrolling. One other website used ``appending infinite scroll'' that adds DOM elements to the website upon scrolling, and \system{} successfully detected all initial and subsequent elements. 

\system{} highlighted task-relevant web elements with an average latency of 10.68s ($\sigma = 6.52s$) and cost of \$0.102 ($\sigma = \$0.08$) per page when tested across 20 websites. Each page contained on an average 3549 text elements ($\sigma = 2561.98$), 79.15 images ($\sigma = 81.78$), and 99.1 SVGs ($\sigma = 85.96$). \system{} filters out less informative text elements with fewer than 3 characters directly on the client side to reduce inference costs and improve responsiveness. This filtering reduced the number of text elements sent to GPT-4o by an average of 79.4\% ($\sigma = 9.24\%$) of the full set. We computed the number of input ($\mu = 7684.95, \sigma = 5288.79$) and output ($\mu = 4235.3, \sigma = 3671.39$) tokens to GPT-4o to calculate cost. The added latency and cost per scroll both increase approximately proportionately with the number of scrolls required. 

We also checked accessibility of these websites to check if and how their accessibility features can affect \system{}'s performance. The average accessibility score of these 20 websites, measured using Google Lighthouse ~\cite{Chrome_for_Developers_2025} was 83.8 ($\sigma = 22.6$). Common issues reported by Lighthouse on these sites included low contrast issues, missing ARIA labels, and unlabeled buttons. \system{} supports any website using standard HTML elements like text, buttons, and SVGs by accessing them through the DOM. It does not support inaccessible content rendered via WebGL or Canvas as they cannot be read or analyzed. None of the 20 websites contained either of these content.

\begin{figure*}
    \centering
    \includegraphics[width=7in]{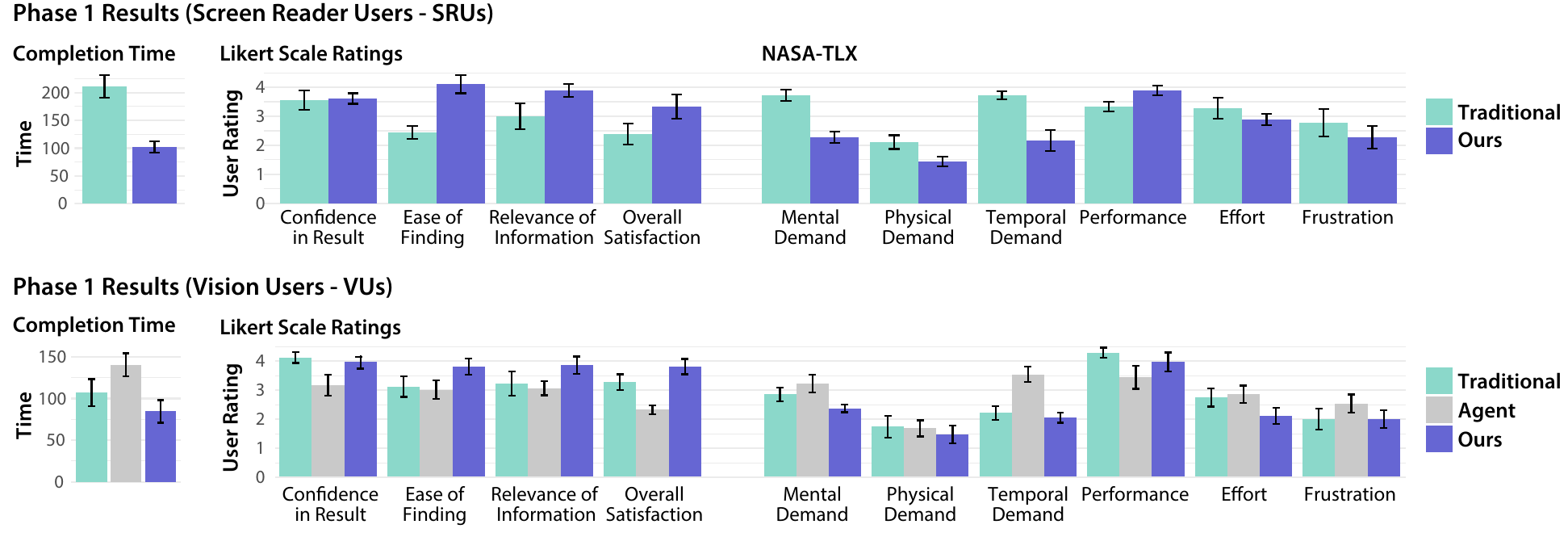}
    \caption{Phase 1 results for SRUs and VUs: The bar plots compare user performance and subjective ratings across three interfaces —Traditional, Agent, and Ours (\system{}). VUs and SRUs completion times and ratings for NASA-TLX (on a scale from 1-low to 5-high), aggregated across all 3 task types (information seeking, navigation, and decision making) for Phase 1.}
    \Description{Two sets of bar charts show the results of Task 1 for Screen Reader Users (SRUs) and Vision Users (VUs), comparing three systems: Traditional, Agent (only in the VU chart), and Ours. Each set includes bar plots for Completion Time, Likert Scale Ratings (Confidence in Result, Ease of Finding, Relevance of Information, Overall Satisfaction), and NASA-TLX metrics (Mental Demand, Physical Demand, Temporal Demand, Performance, Effort, Frustration). In both groups, “Ours” (dark blue bars) generally shows better outcomes—lower task time, higher user satisfaction, and reduced cognitive load—compared to the other conditions. Error bars indicate standard error.}
    \label{fig:task-1-results}
\end{figure*}

\section{User Evaluation}

To assess the impact of \system{} on efficiency and user experience for task-oriented web browsing, we conducted a study with 6 vision users (VUs) and 6 screen reader users (SRUs). The study consisted of a controlled comparison of users completing tasks with a traditional browsing, \system{} (for SRUs and VUs), and a task automation agent (for VUs only\footnote{We cut agents from the SRU study for time as the SRU studies took a long time, agents often encountered errors for VUs, and initial pilots with SRUs demonstrated difficulty working around such errors due to switching between the side bar and main page.}). 
Then, users got to use \system{} for two websites and tasks of their choice and provided open-ended feedback about the benefits, risks, and future use cases of \system{}. 
Through the user study, we aim to answer the following research questions:
\begin{itemize}
    \item[\textbf{R1.}] How does \system{} influence task completion time and perceived effort compared to traditional web browsing? 
    \item[\textbf{R2.}] How does \system{} impact browsing behaviors and experience for different types of tasks? What tasks do users want to use or not use \system{} for in daily browsing?
    \item[\textbf{R3.}] How do users perceive control, trust, and adaptability in our system that guides attention based on task relevance?
\end{itemize}

\begin{table*}[]
\footnotesize
\resizebox{\textwidth}{!}{\begin{tabular}{@{}lrllllll@{}}
\toprule
SRU ID & \multicolumn{1}{l}{Age}                           & Gender & Visual Impairment & Age of Onset & Screen Reader(s)  & Web Expertise  & Frequently Visited Websites                                      \\ \midrule
1      & \multicolumn{1}{l}{46}                            & M      & Totally blind     & Birth        & NVDA, VoiceOver & Expert & Google Search, Claude, Grok, Amazon, PBC, NDTV                               \\
2      & \multicolumn{1}{l}{31}                            & M      & Totally blind     & Birth        & NVDA   & Expert  & Salesforce, Gmail, GoodReads, Amazon, YouTube),   \\
3      & \multicolumn{1}{l}{55}   & F      & Totally blind     & Birth        & NVDA, Jaws & Intermediate  & Library, E-commerce, Stocks websites \\
4      & \multicolumn{1}{l}{34} & M      & Totally blind     & Birth        & NVDA, Jaws & Expert  & News, Information, StackOverflow, Social media\\
5      & \multicolumn{1}{l}{52} & M      & Totally blind     & 25           &JAWS  & Expert  &  Databases, Banking   \\
6      & \multicolumn{1}{l}{63} & F      & Totally blind     & Birth        & JAWS  & Expert  & Work Website                                                   \\ \bottomrule
\end{tabular}}
\caption{Participant details for SRUs in the user evaluation.}
\label{tab:sru-table}
\end{table*}

\begin{table}[ht]
\footnotesize
\normalsize
\centering
\begin{tabularx}{\columnwidth}{@{} p{0.5cm} l l p{1cm} p{4cm}@{}}
\toprule
VU ID & Age & Gender & Web Expertise & Frequently Visited Websites \\
\midrule
1 & 27 & M & Expert & ChatGPT, Google Search, Reddit \\
2 & 27 & F & Expert & Google Scholar, Amazon, Social Media, YouTube, ChatGPT, Github \\
3 & 24 & M & Expert & Social Media, Video Streaming Platforms, Arxiv, Wikipedia, E-commerce \\
4 & 28 & M & Expert & Arxiv, Google Scholar, Quora, Social media, Wikipedia, Amazon \\
5 & 26 & F & Expert & OneNote, iStock, Instagram, LinkedIn, Marketplace, Email, Dating Apps \\
6 & 30 & F & Expert & ChatGPT, Social media, Netflix, Youtube, Google Scholar, Medium \\
\bottomrule
\end{tabularx}
\caption{Participant details for VUs in the user evaluation.}
\label{tab:vu-table}
\end{table}

\subsection{Method}

Our user evaluation included both VUs and SRUs who completed 3 controlled tasks with multiple browsing options (Phase 1), then 2 open-ended tasks on websites of their choice with \system{} (Phase 2). 
All participants were then provided an opportunity for open-ended feedback about risks and benefits of  task-based navigation in a semi-structured interview. The study was an hour and half long, conducted in a 1:1 session via Zoom and in-person, and approved by our institution’s IRB. We compensated participants with \$40 USD for their participation.

\subsubsection{Participants}
We recruited 6 VUs (Table ~\ref{tab:vu-table}) and 6 SRUs (Table ~\ref{tab:sru-table}) who frequently used the web for daily tasks. We recruited VUs using word-of-mouth at our organization and SRUs using internal mailing lists. SRUs used a variety of screen readers (NVDA, Jaws, VoiceOver).

\subsubsection{Materials}
For Phase 1 of the user study, we pre-selected three websites to represent a range of common web use cases and interaction styles: (1) an information-dense website (Wikipedia ~\cite{Wikipedia_2025a}) to simulate information-seeking behavior, (2) a news website (CNN ~\cite{CNN}) that combines text, images, links, and advertisements to represent exploratory browsing and navigation, and (3) an e-commerce website (H-E-B ~\cite{Heb.com}) to reflect goal-oriented decision-making. For each website, we created three task prompts using the template \code{<action> <entity> <constraints>}, similar to prior work ~\cite{shi2017world}. 
This approach enabled us to systematically generate multiple task variants with varying placeholders, but of comparable complexity.
For example, in the task "\textit{Find me the cheapest low sugar vanilla greek yogurt}", the action would be "Find me", entity would be "vanilla greek yogurt", with constraints "low sugar" and "cheapest". 
Exploratory and navigation tasks were based on recent events and required participants to browse from the homepage of a news site to a relevant article. These tasks could not be solved through keyword search or hotkeys (e.g., Ctrl+F) alone and demanded reading and synthesizing content across multiple sections. For instance, one navigation task asked, "\textit{I'm interested in going to Coachella. When is Post Malone's concert?}". This required users to interpret event-related articles and infer the answer based on contextual information. 
Similarly, the information-seeking tasks involved identifying the most relevant section of a dense, text-heavy page that contained the answer to a prompt rather than a simple lookup. 

We used the same pre-processed results for the 9 website-task pairs derived from the 3 structured tasks on the 3 pre-selected websites to maintain consistency across participants. 
The selected websites (CNN, H-E-B, and Wikipedia) support common everyday tasks and are screen reader compatible. Google Lighthouse ~\cite{Chrome_for_Developers_2025} scored the websites 70, 96, and 82 (out of 100 for accessibility), respectively.
See Appendix for the full list of pre-selected websites and tasks used in Phase 1 of the study.

For Phase 2 of the user study, we asked participants to select two websites they frequently used and to identify a common task they typically performed on each. This evaluated \system{} in naturalistic settings reflecting real-world goals. 
Participants' selections spanned social forums, niche e-commerce platforms, personal blogs, and news articles. Some participants chose familiar sites they used often, while others selected websites where they had previously experienced usability challenges.

We ran \system{} in real-time during the study for the VU-selected websites and tasks. They remotely interacted with \system{} using Zoom's Remote Control feature, which allowed them to control the research team’s machine running the extension without local installation. However, for SRUs, remote control access via Zoom was not compatible with screen reader softwares. To address this, we asked SRUs to submit their selected websites and tasks before the study session. We then processed those pairs in advance and shared the \system{} output with SRUs during the study. 
The 24 participant-selected websites had an average Lighthouse accessibility score of 81.25 ($\sigma =$ 12.34), and no participants reported accessibility issues during the study.  

\subsubsection{Procedure}
We first asked participants a series of demographic and background questions about their strategies and challenges while navigating the web. Participants then completed a two-phase study: a controlled pre-selected website-task pair phase (Phase 1), and an open-ended participant-selected website-task pair phase (Phase 2).  

\textbf{Phase 1.} In Phase 1, VUs rated, selected, and provided open feedback on traditional browsing, human-agent navigation, and \system{} for pre-selected 9 website-task pairs. 
We selected an extension of an open-source agent TaxyAI~\cite{taxyai} called CowPilot~\cite{huq2025cowpilot} as our agent option as it allowed users to pause agent execution to navigate around frequent agent errors. 
For the same website-task pairs, SRUs compared and provided feedback on traditional browsing and \system{}. For each browsing method, we provided a short tutorial for the participants on an example website-task pair not included in the study.
Then, for each website-task-pair, we provided the pair and recorded the time the participants took to complete the task. 
We then asked participants to rate on a 5-point Likert scale the NASA-TLX metrics ~\cite{hart1988development}, ease of finding information, relevance of information on webpage to user's task, confidence in task completion, and satisfaction of using the browsing method. 
For VUs we then provided the other two browsing methods for the same website and repeated the rating questions. For SRUs, we then provided the second browsing method for the same website and repeated the rating questions. 
We randomized and counterbalanced the order of the methods of browsing (three methods for VUs and two methods for SRUs) for each participant and across participants for each website-task pair to mitigate ordering effects. After participants provided ratings for every web navigation method for a website, we asked participants to select which navigation method they preferred, and then obtained open-ended feedback.          

\textbf{Phase 2.} In Phase 2, we simulated real-time use of \system{} by asking participants to complete a task on each of two selected websites. Participants selected one task for each of two common web behaviors: a) information seeking (required locating information on a webpage) and b) navigation and exploration. 
For each website-task pair, we asked participants to provide their ratings and feedback on system's support for \system{}, control of browsing experience, helpfulness of different visualization options, trust of the system, on the system's ability to accommodate changes in information needs with task progression.

\textbf{Semi-Structured Interview.} We conducted a semi-structured interview asking participants about their perspectives on the benefits and drawbacks of task-based browsing and about potential future use of \system{}.

\subsubsection{Analysis}
We recorded and transcribed the interviews. To examine participants' feedback on task-specific web browsing, one researcher read interview transcripts to derive themes through affinity mapping. 

\subsection{Results}
\system{} significantly reduced task completion time for SRUs. On average, \textbf{SRUs completed tasks 52\% faster with \system{} compared to traditional browsing method}, with mean task times decreasing from 211 seconds ($\sigma = 49.1$) to 102 seconds ($\sigma = 26.2$), \textit{p} < 0.05. While VUs also saw a decrease in task time from 107 seconds ($\sigma = 39.9$) to 84 seconds ($\sigma = 33.3$), this difference was not statistically significant. Importantly, \textbf{\system{} reduced the performance gap between SRUs and VUs from 2x to 1.2x}. 11 of 12 participants (all 6 VUs and 5 of 6 SRUs) expressed interest in using the system in the future.

\subsubsection{Feedback from SRUs on \system{}}
\textbf{Reduced Cognitive Load, Clutter, and Task Time.}
SRUs appreciated \system{} for its ability to reduce cognitive effort during web navigation. Quantitatively, participants reported lower mental demand when using \system{} ($\mu = 2.28, \sigma = 0.491$) compared to traditional browsing method ($\mu = 3.72, \sigma = 0.491$) on a 5-point scale. Participants consistently emphasized that \system{} could effectively strip away distracting page elements and extraneous content, making relevant information easier to locate and process. For instance, SRU3 remarked, \textit{``It cleans out all of the top part of the web page [menu] that you have to sometimes get through. It is a lot more quicker!''}. For a website-task pair chosen by SRU1 for Phase 2 of the study, SRU1 remarked, \textit{``[\system{}] wins hands down because it lets you cut out a lot of the clutter. It’s much easier to use and you can get things done a lot quicker! You can really narrow down what you need to do''}. \\
\indent \textbf{Utility with Clear and Ambiguous Goals.} 
\system{} supported both focused and exploratory tasks for SRUs (RQ2). Participants expressed that the system was especially effective when participants had a clear task or goal in mind. SRU1 explained, \textit{``I would use task-based [system] when I am looking for very specific answers to questions. Hands down, no questions asked''}. SRU2 and SRU3 emphasized that when the information need was well-defined, the system offered a direct path to relevant content with minimal effort. 
Participants also found \system{} helpful in situations where they lacked an exact search term and also highlighted \system{}’s usefulness in exploratory or fuzzy queries. SRU3 and SRU4 emphasized \system{}'s strength for scanning complex pages, \textit{``It can logically parse a lot of information and extract things I wouldn’t be able to extract otherwise''}. Task Mode can flexibly adapt to a range of user goals and levels of specificity. Participants appreciated being able to pivot goals without restarting the entire interaction, noting that the system could re-prioritize relevant content based on updated goals. \system{} scored highly ($\mu = 4.08, \sigma = 0.79$) for accommodating task changes mid-way through exploration. \\
\indent \textbf{Preservation of Webpage Structure for Simplicity and Control.}
Participants also appreciated that the system retained the high-level layout of webpages, helping in spatial orientation and context. SRU4 remarked, \textit{``It didn’t crop out any search field or something that was relevant. It kept the layout''}. This feature was especially important for users familiar with certain websites. For example, SRU1 who regularly audits their blog for heading structure, stated, \textit{``It is extremely good! I wanted to do a full pulse scan of my personal site to ensure that heading levels are valid and \system{} was able to help me correctly identify the phrase that was not intended for a heading. It was very helpful''}.
\system{} was rated highly for simplicity and control over the browsing experience ($\mu = 4.41, \sigma=0.35$) on a 5-point scale. 
SRUs appreciated that that despite being a filtered view, the system maintained navigational freedom and browsing experience for users. SRU1, SRU3, SRU4, and SRU5 all reported that \system{} was simple to use. SRU3 reflected, \textit{``It did not feel restrictive at all''}. SRUs rated \system{}'s support for task-based navigation favorably ($\mu = 4.5, \sigma=0.79$). \\
\indent \textbf{Tensions with Trust and Context.} While the system was well-received overall, few participants expressed initial hesitations regarding potential loss of context. SRU1 stated, \textit{``The method is sort of unnerving, because I know I’m missing on something''}. However, trust increased with continued use ($\mu = 4.41$, $\sigma = 0.66$). 
SRU4, who initially remarked, \textit{``I work with AI myself, and I know it has the tendency to hallucinate''} later affirmed, \textit{``I quite trust it. It seems to present the right information''}. This is supported by participants (SRU1, SRU3, SRU4, SRU5) who actively cross-checked the task-filtered version with the original page and found key content intact. \\
\indent \textbf{Suggested Improvements.} We observed several opportunities for improvement based on participants' feedback. Participants (SRU4, SRU6) emphasized the importance of displaying complete information blocks such as a heading along with its associated paragraph rather than isolated snippets of relevant text. Sometimes, body text appeared without its associated heading, which led to confusion about the context. SRU1 and SRU4 also pointed out that certain irrelevant elements, such as site navigation links or promotional banners, were not always fully removed. This likely occurred because these elements shared structural similarities (e.g., similar HTML tags or surrounding context) with relevant content, revealing opportunities to further refine threshold calibration or improve semantic filtering in our pipeline. \\
\indent \textbf{Use Cases Identified by SRUs.}
Beyond the structured tasks explored during the study (e.g., online shopping and locating library events), SRUs identified broader applications of task-specific navigation across diverse domains. SRU1 and SRU4 noted its potential in monitoring real-time financial data, where quickly identifying relevant metrics or news is essential for timely decision-making. Others suggested applications in navigating complex policy or healthcare websites, where dense content structures often obscure the specific details users seek. In the context of academic research, participants described how \system{} could help surface pertinent information from lengthy articles or institutional sites. 
SRU4 emphasized the system’s utility  stating, “This was the only way I could quickly find the specific information I was looking for”. These reflections emphasize how task-specific navigation can improve both accessibility and efficiency for SRUs.

\subsubsection{Feedback from VUs on Traditional Browsing and Human-Agent Navigation.} VUs expressed frustration with the human-agent collaborative browsing, particularly when navigating complex websites or switching control between user and agent. Frequent back-and-forth between the system and the user was also seen as undesirable, ~\textit{``I would eventually need to do more frequent takeover. Cause I think taking over is harder, because, like, I don't know what it was up to so far''} added VU2. Although the agent could sometimes direct users to the correct webpage, it lacked reliability in surfacing relevant content, often leading to confusion and reduced trust. VU1 and VU3 commented, \textit{``It was good at finding the right webpage, but it was really bad at highlighting the information within it. I had to do most of the effort myself''}. Trust issues stemmed from both a lack of transparency and unpredictability in the agent's behavior, VU1 highlighted \textit{``The biggest complaint is that is very unreliable as it suddenly stopped working… and it was not super clear, why it is doing [something]''} and VU3 added, \textit{``It's harder to trust the agentic stuff because you don't know… if it's missing something''.}
\subsubsection{Feedback from VUs on \system{}}
\textbf{Eases Information Parsing and Reduces Cognitive Overload.}
Participants emphasized that \system{} reduced the amount of information they had to visually process. 
By surfacing only the most relevant content, the system minimized the need to perform traditional search techniques. VU1 noted, \textit{``I like the fact that it reduced the information I need to pass through. And it was like clearly highlighting what I need to focus [on] first without me manually trying to do control F''}. Participants highlighted \system{}'s ability to quickly direct attention that the interface reduced mental fatigue, particularly when browsing through dense (e.g., Wikipedia) or unfamiliar websites. VU3 and VU5 shared, \textit{``It was less mentally overloading. I don’t think I would like to read everything. This just made it much easier to skim!''}.
Task-specific navigation was seen as a mechanism not only for supporting targeted tasks but also for enhancing general productivity by filtering out irrelevant or distracting content. P2 commented, \textit{``I can also think of it as improving a lot of productivity by filtering out distracting things''}. P3 similarly noted, \textit{``It helps focus, like where you should be looking.''} \\
\indent \textbf{Supports Fine-Grained Control.}
Participants appreciated the system’s ability to facilitate fine-grained exploration of content based on specific task criteria. 
For example, VU1, VU3, and VU4 remarked, \textit{``It acts like a more fine-grained control to look for specific products whose criteria don’t have filters by default on the webpage''}. This selective presentation allowed users to locate task-relevant information that may not have been directly accessible through standard page elements or filters.\\
\indent \textbf{Improves Trust Through Source Referencing and Negative Verification.}
Unlike generative systems that paraphrase or synthesize information, \system{} preserved user trust by showing relevant content directly from its source ($\mu = 4.08, \sigma = 0.88$). VU2 and VU4 shared, \textit{``It is pointing me to a section of a page where I can find relevant information, it's not giving me extra information''}.
Beyond helping users find information, \system{} was also useful for confirming the absence of expected content. VU5 noted, \textit{``The system was also helpful to verify that a particular type of information was not on the page, and I was confident about it''}. This function allowed participants to confidently dismiss irrelevant pages without exhaustive manual inspection. \\
\indent \textbf{Increases User Control with Minimal Learning Curve.}
VUs felt they retained more control over the browsing process, as the system surfaced relevant information while allowing manual oversight ($\mu = 4.25, \sigma = 0.89$). 
Participants also found the system to be intuitive and requiring little effort to adopt, even for first-time users. VU1 remarked, \textit{``It was not too different from traditional browsing… easy enough for me to adapt''} and VU3 added, \textit{``Compared to traditional browsing, it’s nice because it’s not super obtrusive''}.\\ 
\indent \textbf{Captures Semantic Rather Than Surface-Level Relevance.}
Participants noted that the system supported semantically meaningful highlighting even when exact keyword matches were unavailable. VU3 explained \textit{``If the structure of the website just doesn't happen to coincide with the question you're asking you have to pretty much go through everything, this helped with that''}. This semantic alignment was particularly helpful when task goals were vague or used different terminology than the source page.
The system’s use of graded highlighting was appreciated for conveying how strongly each region of a webpage aligned with the task. The system was also able to surface task-relevant information even when it was spread across multiple webpages, which was particularly helpful for multi-step tasks. P5 and P6 noted, \textit{``It got the relevant information right across multiple pages for the task.''} \\
\indent \textbf{Suggested Improvements.} VUs also identified several areas for refinement with respect to \system{}'s usability and interface preferences. First, color visualization preferences varied. While some participants found the use of color gradients helpful for identifying relevant content, others felt that the gradients introduced visual clutter. One suggestion to improve readability was to highlight relevant text using a background fill instead of borders. Second, participants expressed differing opinions on opacity control for irrelevant content. Some preferred complete opacity reduction (i.e., making irrelevant content fully invisible) to minimize distractions and streamline focus. Others favored faint visibility, explaining that even low-relevance content might still contain useful context or keywords. They explained that fully hiding this content forced them to trust the system’s judgment entirely, whereas faint visibility allowed them to independently verify what was being filtered out, providing an added sense of control and confidence during information-seeking tasks. Third, participants requested improved navigation support between highlighted sections. Specifically, they expressed interest in functionality that would allow them to jump between relevant blocks or view a summary of each one. In the current version, users must manually scroll through the entire page to locate task-relevant content.

\begin{table*}[]
\resizebox{\textwidth}{!}{\begin{tabular}{llllllll|llllll}
\toprule
\multicolumn{2}{l}{\multirow{2}{*}{}} & \multicolumn{2}{l}{\makecell{Traditional\\Browsing}} & \multicolumn{2}{l}{\makecell{Task-Specific\\Browsing}} & \multicolumn{2}{l}{} & \multicolumn{2}{l}{\makecell{Human-Agent\\Browsing}} & \multicolumn{2}{l}{\makecell{Task-Specific\\Browsing}} & \multicolumn{2}{l}{} \\
\multicolumn{2}{l}{} & $\mu$ & $\sigma$ & $\mu$ & $\sigma$ & $p$ & $Z$ & $\mu$ & $\sigma$ & $\mu$ & $\sigma$ & $p$ & $Z$ \\
\midrule
\midrule
\multicolumn{1}{c}{\multirow{6}{*}{NASA-TLX}} & Mental demand & 2.94 & 0.61 & 2.44 & 0.34 & 0.281 & -1.26 & 3.33 & 0.76 & 2.44 & 0.34 & 0.058 & 1.99 \\
\multicolumn{1}{c}{} & Physical demand & 1.78 & 0.96 & 1.50 & 0.78 & 0.789 & -0.57 & 1.72 & 0.71 & 1.50 & 0.78 & 0.357 & 1.08 \\
\multicolumn{1}{c}{} & Temporal demand & 2.28 & 0.61 & 2.11 & 0.46 & 0.855 & -0.22 & 3.67 & 0.67 & 2.11 & 0.46 & \textbf{0.031} & 2.20 \\
\multicolumn{1}{c}{} & Performance & 4.44 & 0.46 & 4.11 & 0.83 & 0.371 & -1.41 & 3.56 & 1.00 & 4.11 & 0.83 & 0.418 & -0.84 \\
\multicolumn{1}{c}{} & Effort & 2.83 & 0.81 & 2.17 & 0.72 & 0.281 & -1.26 & 2.94 & 0.77 & 2.17 & 0.72 & 0.106 & 1.69 \\
\multicolumn{1}{c}{} & Frustration & 2.06 & 0.90 & 2.06 & 0.80 & 1.000 & 0.00 & 2.61 & 0.80 & 2.06 & 0.80 & 0.281 & 1.05 \\
\multicolumn{2}{l}{Confidence in Result} & 4.28 & 0.49 & 4.11 & 0.58 & 0.789 & -0.57 & 3.28 & 0.93 & 4.11 & 0.58 & 0.100 & -1.94 \\
\multicolumn{2}{l}{Ease of Finding} & 3.22 & 0.91 & 3.94 & 0.71 & 0.181 & 1.71 & 3.11 & 0.83 & 3.94 & 0.71 & 0.106 & -1.69 \\
\multicolumn{2}{l}{Relevance of Information} & 3.33 & 1.07 & 4.00 & 0.76 & 0.170 & 1.49 & 3.17 & 0.62 & 4.00 & 0.76 & 0.071 & -1.91 \\
\multicolumn{2}{l}{Overall Satisfaction} & 3.39 & 0.68 & 3.94 & 0.68 & 0.410 & 1.07 & 2.39 & 0.39 & 3.94 & 0.68 & \textbf{0.035} & -2.21 \\
\multicolumn{2}{l}{Completion Time (seconds)} & 106.87 & 39.91 & 84.43 & 33.33 & 0.563 & -0.73 & 140.55 & 33.48 & 84.43 & 33.33 & 0.156 & 1.57 \\
\bottomrule
\end{tabular}}
\caption{VUs: Comparison of Traditional, Agent, and Task-Specific Browsing for sighted users. Task-Specific Browsing significantly reduced temporal demand and improved overall satisfaction compared to other interfaces. Metrics include NASA-TLX scores, user ratings, and completion time with mean $(\mu)$, standard deviation $(\sigma)$, and Wilcoxon test results $(p, Z)$.}
\end{table*}

\begin{table*}[]
\begin{tabular}{llllllll}
\toprule
\multicolumn{2}{l}{\multirow{2}{*}{}} & \multicolumn{2}{l}{\makecell{Traditional\\Browsing}} & \multicolumn{2}{l}{\makecell{Task-Specific\\Browsing}} & \multicolumn{2}{l}{} \\
\multicolumn{2}{l}{} & $\mu$ & $\sigma$ & $\mu$ & $\sigma$ & $p$ & $Z$ \\
\midrule
\midrule
\multirow{6}{*}{NASA-TLX} 
& Mental demand      & 3.72 & 0.49 & 2.28 & 0.49 & \textbf{0.034} & -2.23 \\
& Physical demand    & 2.11 & 0.58 & 1.44 & 0.40 & \textbf{0.035} & -2.21 \\
& Temporal demand    & 3.72 & 0.33 & 2.17 & 0.86 & 0.059 & -2.11 \\
& Performance        & 3.33 & 0.42 & 3.89 & 0.40 & 0.054 & 2.13 \\
& Effort            & 3.28 & 0.88 & 2.89 & 0.46 & 0.461 & -0.97 \\
& Frustration       & 2.78 & 1.17 & 2.28 & 0.95 & 0.100 & -1.94 \\
\multicolumn{2}{l}{Confidence in Result} 
                    & 3.56 & 0.81 & 3.61 & 0.44 & 0.855 & 0.65 \\
\multicolumn{2}{l}{Ease of Finding} 
                    & 2.44 & 0.54 & 4.11 & 0.78 & \textbf{0.031} & 2.20 \\
\multicolumn{2}{l}{Relevance of Information} 
                    & 3.00 & 1.07 & 3.89 & 0.54 & 0.156 & 1.57 \\
\multicolumn{2}{l}{Overall Satisfaction} 
                    & 2.39 & 0.90 & 3.33 & 1.03 & 0.058 & 2.00 \\
\multicolumn{2}{l}{Completion Time (seconds)} 
                    & 211.25 & 49.12 & 102.21 & 26.18 & \textbf{0.031} & -2.20 \\
\bottomrule
\end{tabular}
\caption{SRUs: Comparison of Traditional and Task-Specific Browsing for blind users. Task-Specific Browsing significantly reduced mental and physical demand, and improved ease of finding and completion time. Table reports mean $(\mu)$, standard deviation $(\sigma)$, and Wilcoxon test results $(p, Z)$ for each metric.}
\end{table*}

\section{Discussion}
Our findings confirm prior work showing that surfacing task-relevant content (\textbf{DG1}) improves SRU efficiency ~\cite{giraud2018web}. In contrast, \system{} can be automated at scale using LLMs, applied to real-world websites, and 
can significantly improve web navigation efficiency for screen reader users (SRUs), while maintaining performance for vision users (VUs), such that the performance gap between SRUs and VUs is reduced (\textbf{RQ1}).
\system{} keeps users in control unlike systems that hide fixed distractions ~\cite{Apple2024DistractionControl, lee2025purpose} or automate actions ~\cite{lu2024weblinx, zheng2024webolympus}. Participants valued its transparency and ability to review content in context.
11 of 12 participants wanted to use \system{} in the future and a key advantage highlighted by SRUs and VUs in qualitative feedback was its capacity to reduce effort while searching for content of interest (\textbf{RQ1}). 
To make browsing as easy as possible, our system preserved webpage structure (\textbf{DG2}) and participants reported that this made the interface simple to use (SRUs) and promoted trust in the application compared to automated agents (VUs). 
Our prototype maintained persistent task context across navigation (\textbf{DG3}) and supported real-time task updates as users' goals evolved (\textbf{DG4}). 
Participants found this feature to be particularly useful in the open-ended task as participants needed to formulate their own task and typically visited multiple pages. 
After using \system{} for our assigned tasks and their own use, participants reported utility of \system{} for both well-defined queries and exploratory searches where exact query terms are unknown. Participants shared that they expected \system{} to be most useful in the future for complex information-seeking tasks, navigation on unfamiliar websites, and filtering content on dense information pages (\textbf{RQ2}). 
\system{} does not replace or override traditional screen reader interactions such as navigating by heading levels or landmarks. Instead, it complements these existing strategies by prioritizing what content is made available to the user, helping SRUs locate relevant information faster within the familiar interaction paradigm. Participants reported continuing to use screen reader shortcuts like heading navigation in \system{}, but with less need for exhaustive scanning, since irrelevant regions were minimized. Similarly, VUs still scrolled and scanned visually but spent less effort discerning what to focus on.
\system{} offered multiple visualization modes (\textbf{DG5} - support personalization) to address the diverse needs of both VUs and SRUs. Our study revealed different preferences among users - SRUs preferred complete filtering of irrelevant content, while VUs expressed varied preferences for visualization methods (color gradients, opacity reduction, or complete filtering) (\textbf{RQ3}).
Although this mode was primarily designed for VUs and SRUs, it can also benefit other user groups. For instance, low-vision users may prefer reduced visual complexity, users with motor impairments may find fewer interactive elements easier to navigate, and users with attention challenges (e.g., ADHD) may benefit from a decluttered interface. Additionally, users can dynamically adjust the threshold to control content displayed, offering flexibility based on individual preferences and task complexity.
Our work demonstrates how designing new interactions simultaneously for visual and non-visual access can reduce rather than reinforce accessibility disparities in web navigation. \\

\subsection{Support for Inclusive and Universal Design}
\system{} demonstrates how systems can be designed to benefit a broad range of users while also supporting specific access needs. \system{} serves as a \textit{universal design} solution that improves navigation efficiency for all users, regardless of ability by prioritizing task-relevant content and structured navigation. At the same time, it incorporates \textit{inclusive design} by offering personalized views tailored to different user preferences and access strategies. Most new AI interactions (e.g., agents, image tools) are designed for sighted users first, then retrofitted for BLV users. This retrofitting often fails to address fundamental interaction mismatches. For example, agents that rely on mouse movements or visual layout cues are inherently misaligned with non-visual workflows. Their low task success rates (often below 12\%) ~\cite{xie2024osworld} highlight their unreliability as standalone solutions for accessibility. Contrary to assumptions that such agents will replace traditional interaction methods, our findings show that SRUs continue to rely on established screen reader strategies (e.g., heading and landmark navigation). \system{} complements these workflows rather than replacing them. By centering accessibility from the start, it avoids the pitfalls of retrofit design and demonstrates how inclusive systems can reduce, rather than reinforce accessibility gaps. Our findings show the promise of designing general-purpose tools for SRUs first (universal), then adapting them for VUs (inclusive).

\subsection{Limitations and Future Work}
Our work introduces several areas for future work: 

\subsubsection{Improving \system{}'s Performance}
While our system supported users performing our controlled tasks and their selected tasks, it occasionally produced errors. 

\textbf{Performance of Task-Relevance Element Scoring. } 
Using LLMs to assign task relevance scores introduces risks since their output can be subjective and inconsistent. While LLMs are good at matching content to the task prompt, they can miss important interface cues, like buttons or labels, or misunderstand what the user really needs. This can lead to errors where useful elements are not scored as relevant. 
These problems were occasionally worsened by our bottom-up approach, which scored individual words or elements first, and then built up to larger sections. For example, short texts like links within paragraphs received low scores even though the full paragraph or container was relevant. This caused fragmented highlights, where only parts of a relevant block were shown. SRUs reported that while such omissions rarely prevented them from completing tasks, they sometimes created doubt about whether something important was removed. To verify this, some users lowered the relevance threshold or scanned the page at multiple settings. In the future, we could improve performance by adding top-down cues like visual structure or semantic grouping, and allow the system to adapt thresholds based on user needs, especially when high-recall filtering for sensitive tasks is important.

\textbf{Performance of Element Extraction. } 
\system{} can detect and filter out inaccessible content rendered using WebGL or Canvas, but it cannot analyze this content because it lacks semantic structure in the DOM. Even for accessible elements, \system{} may miss content that is dynamically rendered, such as items loaded through infinite scrolling or updated via frameworks like React ~\cite{React_Blog_RSS}. Since these changes don’t always update the DOM immediately, \system{} only captures them if the user manually re-runs the system. We plan to integrate \code{MutationObservers} to track real-time DOM changes and automatically reprocess content as it appears. \\

\subsubsection{Extending \system{} to New Domains} Future research could extend \system{}’s strengths beyond the browser by adapting its core contribution (highlighting task-specific content) to new environments. One direction is adapting \system{} for multimedia content. For instance, \system{} could be adapted to help users navigate videos more easily by automatically highlighting segments relevant to a given task, such as detecting sections of social interaction for behavioral coding. Similarly, when working with audio, \system{} could lower the volume of irrelevant segments, making it easier to skim linearly and focus on key content.
Another opportunity lies in data analysis workflows using task-guided visualizations. Chart elements, filters, or data ranges that align most with the user’s goal could be emphasized, supporting fine-grained control and improved sensemaking in complex datasets. 
Another area of using \system{} is Augmented Reality in physical environments, where it could be used to guide attention towards physical objects relevant to the task at hand. This kind of contextual support is especially helpful for people with low vision, prior research shows that directing attention to relevant items significantly improves task success rates ~\cite{zhao2019designing, zhao2019seeingvr}. A third direction is in managing multi-step workflows, such as tutorials or guided tasks, where \system{} could track progress and highlight interface elements relevant to the current step ~\cite{mysore2018porta}. In all these domains, users often seek help to stay focused and minimize distractions during complex tasks. Prior research shows that users benefit from self-control interventions that reduce distraction and improve task productivity ~\cite{kovacs2021not}, which \system{} could support.

\section{Conclusion}

We present a task-specific browsing system that dynamically filters and highlights webpage content based on the user’s stated goal, aiming to reduce navigation time and cognitive load for both vision users (VUs) and screen reader users (SRUs). We evaluated our system through a combination of technical metrics and a user study with 12 participants (6 VUs, 6 SRUs). Participants reported that the task-specific views surfaced more relevant information, reduced unnecessary exploration, and made it easier to complete web-based tasks. 
We aim to support future research on task-driven web browsing experiences.

\bibliographystyle{ACM-Reference-Format}
\bibliography{references}

\appendix

\section{System Pipeline Prompts}
\subsection{Breakdown of User's Task}

\noindent Given the user's task: \texttt{\{task\}}, break it down into the following components to help identify and prioritize relevant page elements:

\begin{enumerate}
    \item entity: the main object or item the user is interested in
    \item constraints: broad categories of requirements or preferences (e.g., cost, color, brand, filter categories)
    \item actions: logical steps the user might take on a webpage (e.g., filter, compare, add to cart)
    \item default: elements the user expects to see on a webpage (e.g., product name, price, image)
    \item fallback: alternative elements the user can use if primary ones are missing (e.g., search bar, contact support)
\end{enumerate}

\noindent Return in the following JSON format:
\begin{verbatim}
{
    "entity": "...",
    "constraints": ["...", "..."],
    "actions": ["...", "..."],
    "default": ["...", "..."],
    "fallback": ["...", "..."]
}
\end{verbatim}

\subsection{Ranking Task-Specific Text Elements}

User Task: \texttt{\{task\}} \\
Task Steps: \texttt{\{taskBreakdown\}}
\\

\noindent You are interpreting the webpage from a list of text elements to determine which are relevant for completing the user’s task. Do not evaluate elements in isolation. Instead, consider their surrounding elements and whether they belong to a structured unit or repeated group (e.g., a product card, listing, or section).
\\

\noindent If one element in a unit strongly matches the task (e.g., product title, price, size), score associated elements (e.g., “Add to cart”) accordingly based on how much they contribute to completing the task.
\\

\noindent Score each element on a scale from \texttt{0} (irrelevant) to \texttt{100} (critical to the task).
\\

\noindent Key Guidelines:
\begin{enumerate}
    \item Interpret structured groups (cards) from the tags of texts. Treat these as semantic units. Score inner elements similarly only if the card as a whole is relevant.
    \item Match all task constraints (e.g., flavor, price, size, pickup availability). If 2–3 constraints are satisfied, the card is still relevant.
    \item Score repeated actions (like “Add to cart” or “Add to list”) only if tied to a relevant card.
    \item Penalize loose, standalone text (e.g., titles, prices, buttons) not clearly part of a relevant group.
    \item If missing the element would prevent task success, assign a high score.
\end{enumerate}

\noindent Infer relationships based on text proximity, matching patterns, and card repetition structure.
\\

\noindent  Return only an array with \texttt{\{len(batch)\}} scores. No explanation. \\
\noindent Example format: \texttt{[12, 87, 34, 0, 75]} \\
\noindent Text elements to score: \texttt{\{numbered\_elements\}}

\subsection{Ranking Task-Specific Image Elements}
\noindent Given the user's task: \texttt{\{task\}} and the task steps: \texttt{{taskBreakdown}}, evaluate the relevance of each image's alt text in helping the user complete their task. \\

\noindent Score each image description (alt text) from 0 (irrelevant) to 100 (critical to the task). Return only an array of scores. \\

\noindent Example output: \texttt{[12, 87, 34, 0, 75]} \\
\noindent Alt texts to score: \texttt{\{numbered\_alts\}}

\subsection{Identifying SVGs from \texttt{path}}
\noindent Given the following list of SVG paths, identify what each icon represents. For each path, return a short name like "search", "close", "menu", "user", etc.\\

\noindent DO NOT add any other text. Only return a list of \texttt{len(paths)} names in this format: \texttt{["search", "user", "menu", ...]} \\

\noindent SVG paths: \texttt{\{numbered\_paths\}}

\subsection{Ranking Task-Specific SVG Elements}
\noindent Given the user's task: \texttt{\{task\}} and the task steps: \texttt{taskBreakdown}, you are interpreting the context of the current webpage from the icon titles given to you to understand which steps are meaningful and actionable from the breakdown of the user’s task. \\

\noindent Score each icon title on a scale from 0 (irrelevant) to 100 (critical to the task). \\

\noindent Prioritize scoring task-relevant icons higher, as the user must see them to successfully complete their task. Missing these icons would result in a poor experience, which this system aims to prevent. \\

\noindent Return only an array with \texttt{len(batch)} scores. Do NOT include any explanation, notes, or additional text.

\noindent Example output format: \texttt{[12, 87, 34, 0, 75]}\\
\noindent Icons to score: \texttt{numbered\_elements}

\begin{figure*}
    \centering
    \includegraphics[width=7in]{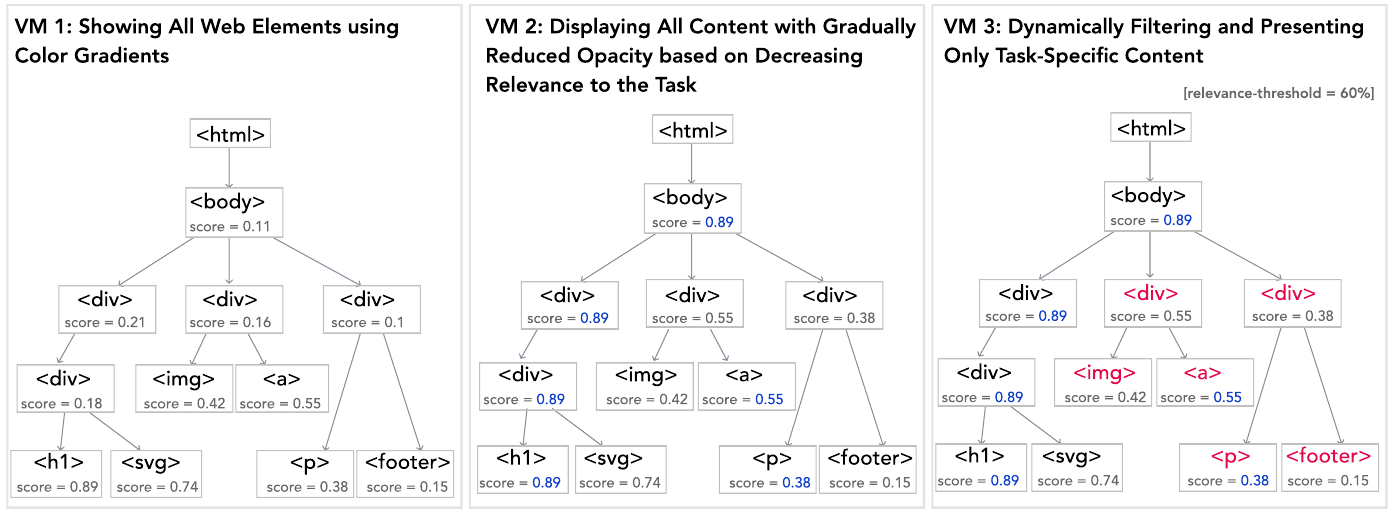}
    \caption{This figure illustrates how task relevance scores are assigned and propagated across the DOM tree in three visualization modes (VM). Each tree represents the DOM hierarchy of a web page, where each node corresponds to a web element (e.g., \code{<div>}, \code{<img>}, \code{<p>}), and edges represent parent-child relationships. Left: Color gradients visualize task relevance per element independently, without propagating scores across the DOM. Leaf nodes are emphasized, while structural containers are visually de-emphasized. Center: Scores are propagated upward (in blue) so parent elements inherit higher child scores. Element opacity is then adjusted to visually fade out less relevant content while preserving layout. Right: A relevance threshold (e.g., 60\%) is applied. Low-relevance nodes (in red) are hidden. Scores are propagated upward to retain necessary ancestors for a focused, task-specific view.}
    \Description{A side-by-side comparison of three DOM tree diagrams illustrating different relevance-based content visualization modes for web pages. Each node represents a web element with an associated task relevance score. Left panel showing all web elements using color gradients: Scores are assigned independently, with no propagation. All elements are shown. Center panel displaying all content with gradually reduced opacity based on decreasing relevance to the task: Scores are propagated upward from child to parent nodes (shown in blue); all content is shown with opacity adjusted based on relevance. Right panel with dynamically filtering and presenting only task-specific content: A relevance threshold of 60\% is applied; low-scoring elements (shown in red) are removed, and scores are propagated upward to retain ancestors of relevant content. Blue text indicates propagated scores. Red text indicates removed nodes.}
    \label{fig:relevance-propagation}
\end{figure*}

\begin{table*}[]
\centering 
\normalsize
\begin{tabular}{@{}llp{2cm}p{9cm}@{}}
\toprule
Task Type                                   & Website                    & \# of Webpages Visited & Pre-selected User Tasks                                                                                        \\ \midrule
\multirow{3}{*}{Information Seeking}        & \multirow{3}{*}{Wikipedia} &               1                      & "How might a cyclone evolve as it moves across a wide range of latitudes?"                                     \\
                                            &                            &              1                      & "Which specific membrane within a plant cell contains the proteins responsible for photosynthesis?"            \\
                                            &                            &              1                      & "What kind of preparations did the athletes undergo to train for the ancient Olympic Games?"                   \\
\multirow{3}{*}{Navigation and Exploration} & \multirow{3}{*}{CNN News}  &               2                      & "Has the stock market improved? How has this impacted gold prices?"                                            \\
                                            &                            &              2                      & "Show me the latest news on the Hudson river crash. How many people were injured?"                             \\
                                            &                            &              2                      & "I’m interested in going to Coachella this year. When is Post Malone’s concert?"                               \\
\multirow{3}{*}{Decision Making}            & \multirow{3}{*}{H-E-B}     &               1                      & "I want to buy the cheapest 4 pack low sugar vanilla greek yogurt."                                            \\
                                            &                            &              1                      & "I want to buy most expensive chocolate ice-cream option that is 1 pint and is available for curbside pickup." \\
                                            &                            &              1                      & "I want to buy a brownie mix that is at least 18 oz under \$4.                                                 \\ \cmidrule(l){1-4} 
\end{tabular}
\caption{Pre-selected Tasks in Part A of User Study}
\label{tab:Pre-Selected Tasks}
\end{table*}

\begin{table*}[]
\centering 
\normalsize
\begin{tabular}{@{}p{1cm}p{3cm}p{2cm}p{10cm}@{}}
\toprule
PID   & Website              & \# of Webpages Visited & Participant-Selected Tasks                                                                                                    \\ \midrule
VU 1  & Reddit r/4kTV        & 3                      & "Show me the reviews on OLED TVs"                                                                                             \\
      & Amazon               & 3                      & "Find me a board game with atleast 1000 ratings and 4.5 stars"                                                                \\
VU 2  & AllTrails            & 2                      & "Show me all hikes within 30 miles of distance"                                                                               \\
      & Personal Website     & 1                      & "What were the publications from recent internship?"                                                                          \\
VU 3  & HyperPhysics         & 2                      & "What is the mistake predicted by Classical Mechanics?"                                                                       \\
      & Wikipedia - Penguin  & 2                      & "How endangered is the smallest penguin?"                                                                                     \\
VU 4  & IoPS                 & 2                      & "Show me latest IMO contest problems"                                                                                         \\
      & Gardeners            & 2                      & "How can I fertilize my garden?"                                                                                              \\
VU 5  & Wikipedia - Tesla    & 1                      & "What are Tesla's latest energy products?"                                                                                    \\
      & Anavrin              & 2                      & "I want to buy the cheapest book nook under \$100"                                                                            \\
VU 6  & Reddit               & 2                      & "What cafes are near UT Austin?"                                                                                              \\
      & Wikipedia            & 1                      & "When did women start playing basketball?"                                                                                    \\
SRU 1 & Personal Website     & 1                      & "Find the text on this page that is a heading but should not be, for better understandability and readability of this article" \\
      & BBC News             & 2                      & "Show me latest news on India"                                                                                                \\
SRU 2 & Sip and Feast        & 1                      & "Show me pasta recipes with shrimps"                                                                                          \\
      & GoodReads            & 2                      & "What are the reviews and ratings of this book?"                                                                              \\
SRU 3  & Arapahoe Libraries   & 2                      & "What are the current events in this library?""                                                                               \\
      & Zack's Stocks        & 1                      & "Show me the rankings for bear and bull stocks for today"                                                                     \\
SRU 4 & Dailymotion - Delish & 2                      & "Show me the first asparagus recipe on this page"                                                                             \\
      & Blog Bit AI          & 1                      & "Show me the first recommended blog platform on this page"                                                                    \\
SRU 5 & Fox News             & 1                      & "Is there any news on the impact of AI jobs?"                                                                                 \\
      & Bloomberg            & 2                      & "Show me the latest technology news"                                                                                          \\
SRU 6 & Educause             & 2                      & "When is the deadline for submitting a conference workshop proposal?"                                                         \\
      & Fairfield            & 1                      & "Who is on the staff of the Center for Social Impact?"                                                                       \\ \cmidrule(l){1-4} 
\end{tabular}
\caption{Websites and Tasks chosen by VUs and SRUs in Part B of User Study}
\label{tab:Participant-Selected Tasks}
\end{table*}

\begin{table*}[]
\centering
\normalsize
\begin{tabular}{@{}lllll@{}}
\toprule
& \multicolumn{2}{l}{SRUs} & \multicolumn{2}{l}{VUs} \\ 
& $\mu$           & $\sigma$          & $\mu$          & $\sigma$          \\ \midrule
Support for task-based navigation       & 4.5         & 0.79       & 4.16       & 0.81       \\
User control of browsing experience     & 4.08        & 1.08       & 4.25       & 0.89       \\
Helpfulness of different visualizations & 3.91        & 1.16       & 4.04       & 0.90       \\
Trust on system                         & 4.41        & 0.66       & 4.08       & 0.88       \\
Accommodation for task changes          & 4.08        & 0.79       & 3.95       & 0.85       \\ \bottomrule
\end{tabular}
\caption{SRUs and VUs ratings for \system{} in Phase 2. \system{} effectively supports both VUs and SRUs through task-specific navigation, enabling users to feel more in control of their browsing experience with higher trust. Table presents mean ($\mu$) and standard deviation ($\sigma$) of each metric.}
\label{tab:my-table}
\end{table*}

\end{document}